\documentclass[preprint,12pt,longtitle,authoryear]{elsarticle}
\usepackage{amsfonts}
\usepackage{amssymb}
\usepackage{amsmath}
\usepackage{graphicx}
\usepackage{epsfig}
\usepackage{color}

\newcommand{\A}{\mathcal A}

\newcommand{\cS}{\mathcal S}

\newcommand{\RR}{{\mathbb R}}

\newtheorem{proposition}{Proposition}

\newcommand{\blue}{\textcolor{black}}

\begin{document}

\title{Modelling aspects of consciousness: a topological perspective}
\author{Mike Steel}
\date{December 2, 2020}

\begin{abstract}
Attention Schema Theory (AST)  is a recent proposal to provide a scientific explanation for  the basis of subjective awareness.  
\blue{In AST, the brain constructs a representation of attention taking place in its own (and others') mind  (`the attention schema').  Moreover, this representation  is incomplete for efficiency reasons. }
This inherent incompleteness of the attention schema results in the inability of humans to understand how their own subjective awareness arises (related to the so-called `hard problem' of consciousness).  \blue{Given this theory, the present paper asks whether a mind (either human or machine-based)  that incorporates attention, and that contains a representation of its own attention, can ever have a complete representation}. Using a simple yet general model and a mathematical argument based on classical topology, we
show that a complete representation of attention is not possible, since it cannot faithfully represent streams of attention. \blue{In this way, the study supports one of the core aspects of AST, that the brain's representation of its own attention is necessarily incomplete.}   \end{abstract}

%  In AST, the brain constructs a representation of attention. That representation is incomplete. All representations in the brain are incomplete, for efficiency. Based on the incomplete information in the brain's representation of attention, people believe they have a mysterious essence inside of them, consciousness, instead of the physically real process of attention that they actually have. 
\address{Biomathematics Research Centre, University of Canterbury, Christchurch, New Zealand}

\maketitle

\bigskip

\noindent{\em Email:} mike.steel@canterbury.ac.nz

\bigskip


\newpage
\section{Introduction}

The question of explaining the basis and nature of human consciousness remains an outstanding scientific question that straddles several disciplines, from neuroscience, cognitive psychology and biophysics \citep{ bar, deh, lau, lin, rol}  to philosophy \citep{cha, den}.   The question is often divided into two parts:  the so-called `easy' and `hard' problems \citep{cha}.  Roughly speaking, the `easy' problem seeks to identify the neural processes that give rise to the experiences one typically associates with consciousness (perception, subjective awareness, introspection, thought and emotion).   The second (`hard') problem is more fundamental:  it asks how is it that humans experience subjective awareness at all (i.e. the experience of `qualia'), rather than simply processing information in an unconscious  but purely computer-like fashion. 

A great deal has been learned in recent decades concerning the first problem, and  theories have been proposed to describe how conscious experience might arise within a brain or  within a sufficiently complex neural network. Two leading approaches are (neuronal) Global Workspace Theory (GWT)  \citep{bar, wal} and Integrated Information Theory  \citep{est, ton}, though many other related ideas have been put forward, including, for example, thalamocortial resonsance \citep{lin} or the emergence of consciousness as a phase transition in brain dynamics \citep{wer}.

Here, we consider an approach to consciousness that deals with both problems, namely Attention Schema Theory (AST), pioneered by Michael Graziano and colleagues \citep{gra, gra2, gra3, web}.  AST builds on  approaches such as GWT  to describe how attention arises within a brain by well-studied processes in which the focus of attention (which can be on external stimuli or internally on  memories or thoughts) is governed by  a competitive process of excitation, reinforcement and inhibition.  AST further suggests that conscious awareness requires an additional aspect -- an internal concept of the self, and an internal cognitive model of how attention works (the `Attention Schema') -- the latter having evolved by natural selection in higher animals, because of the selective advantage of the brain controlling its own attention processes, as well as providing a way for an individual to model the minds of other individuals in social settings \citep{gra}. 

According to AST, a central reason why the `hard' problem of consciousness appears hard is  simply because the attention schema is highly incomplete, because it is a grossly oversimplified internal model that does not include all the neural details of how attention actually arises (greater completeness and additional neural detail in an attention schema would reduce computational efficiency, and so be unlikely to evolve).  \blue{This incompleteness leads individuals to conclude  they have a mysterious essence inside of them, consciousness, instead of the physically real process of attention that they actually have.}

Thus, AST provides a specific structural process for consciousness to arise and for a mind to know that it is conscious, yet without having the ability to completely understand why.  AST also suggests that current machines are unlikely to be conscious but could be designed to be so (though it is not enough to simply increase computational power).   For further details, the reader is referred to \cite{gra, gra2, gra3, web}.

In this paper, we  provide a simple mathematical analysis of a question that arises from AST.  We ask whether a mind could, in principle,  have an attention schema that is complete, in the sense that any stream of attention -- at least over a short time-frame -- can be faithfully represented within the attention schema.    Using a simple argument based on standard concepts from topology, we show that  \blue{a complete attention schema is inherently impossible, providing a further and more fundamental reason for the incompleteness assumption in AST.}

\section{Mathematical model}
The model we describe is particularly simplified; however, by making minimal assumptions and working in  a high-level setting, the results we describe  --- while less detailed --- are more generic than those that might be derived from a detailed and finely-tuned  model. Our approach is based on concepts and results from topology, which is the mathematical study of `space' (and maps between spaces) in a very general sense (topological spaces include many familiar examples that form the basis of physics, biology and other fields in science) \citep{ghi, hat}.

We  view attention within an individual mind as moving across some open, bounded and connected subspace $\A$ of a high-dimensional Euclidean space (thus  $\A$ comprises a $n$-dimensional manifold, with each point in $\A$ having  an open neighbourhood that is (homeomorphic to)  an open $n$-cell).  The space $\A$ could, for example, describe  neuronal firing patterns,  fluctuating neural states and changing parameters within the brain; alternatively, $\A$ might be taken to be a purely phenomenological space, or perhaps some space that involves aspects of both types.  In this paper, we impose no restriction on $n$; for example, $n$ could be  in the order of the number of synaptic junctions in a brain or greater.  

Throughout we will adopt a convention in topology of using  the word  {\em map}  to refer to any continuous function; also, a map is said to be {\em injective} if it is one-to-one. 
 
 We will let $o_t$ denote an open set in $\A$ that is the focus of attention at time $t$ (thus, $o_t$ is also a manifold of dimension $n$). Consistent with approaches such as GWT, we will describe $o_t$ as the region of $\A$  that has reached the threshold required for attention at the given instant $t$. More formally, consider  a function $F$ that assigns a non-negative real value to each pair  $(t,x)$ ($t=$time, $x$ is an element of $\A$); we will assume throughout that $F$  is continuous in both variables and refer to it as 
 an {\em attentional activity function}.

 Given a threshold $\delta>0$, we will formally define $o_t$ to be  the  portion of $\A$ that lies above this threshold at time $t$  and thereby becomes the focus of attention at time $t$.\footnote{One could further allow $\delta$ to  depend continuously on $t$, but this gives no further generality since $\delta$ can be absorbed into  $F$; indeed, by rescaling, one could assume that $\delta =1$.}   More precisely: \begin{equation}
 \label{Feq}
o_t = \{x \in \A: F(t, x) > \delta\}.
\end{equation} 
Notice that $F$ may vary with $t$ while $o_t$ remains constant.\footnote{In topological terms, $o_t$ need not be connected (we give an example shortly), or even contractable,  and $o_t$  could also contain  `holes' of varying dimensions (i.e. non-trivial homology).}

\begin{figure}[h]
\centering
\includegraphics[scale=0.35]{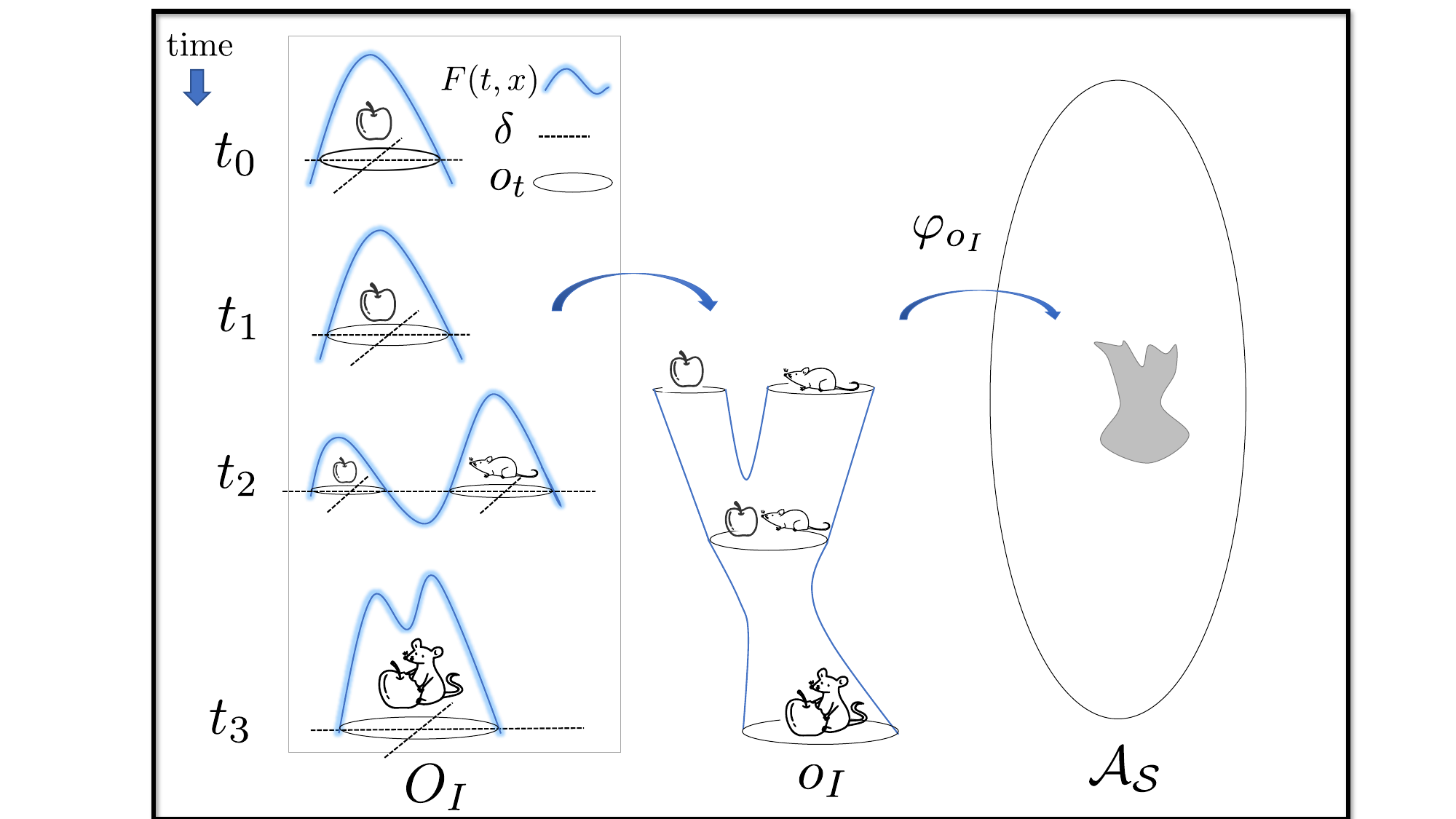}
\caption{A schematic illustration of the model in a highly simplified setting where $o_t$ is two-dimensional.  {\em Left:} A stream of attention $o_t$  of an individual mind described by the portion of $\A$ that has attentional activity (indicated by $F$) larger than a threshold ($\delta$). Initially, attention is focused on the apple but then shifts when a mouse appears, and eventually eats the apple.  The open sets $o_t$ are two-dimensional (in the horizontal plane) and involve a frozen stream of attention on the interval $[t_0, t_1]$, after which a jump occurs (when the mouse is noticed),  resulting in $o_t$ briefly having two components. Subsequently, $o_t$ contains two (or more) moving streams of attention as the mouse moves towards the apple and then begins to eat it. {\em Middle:} The  streams of attention in $O_I$ map to the quotient space $o_I$ (i.e. the initial frozen stream of attention has been collapsed). {\em Right:} A mapping of $o_I$ into a formal attention schema $\A_{\cS}$. Since $\A_{\cS}$ has the same dimension as $\A$ in this example and $o_t$ contains a moving stream of attention, the map cannot be injective (by Proposition~\ref{mainthm}(i)).}
\label{fig1}
\end{figure}

We now introduce two further (related) definitions. First, given an interval $I$ of strictly positive length,  we say that $o_t$  is a  {\em frozen stream of attention} if, for all $t, t' \in I$ with $t<t'$, we have  $o_t = o_{t'}$.  Thus a frozen stream of attention is one where the focus of attention does not shift (as in Fig.~\ref{fig1} between  $t_0$ and $t_1$).  Next, given an (open or closed) interval $I$, let $O_I$ denote the subspace of $I \times \A$ defined by: $$O_I=\{(t, x): t \in I, x \in o_t\},$$ 
and consider the quotient space:  $$o_I = O_I/\sim,$$
where $\sim$ is the equivalence relation on $O_I$ defined by $(t, x) \sim (t', x')$ if $x=x'$ and $o_t$ is a frozen stream of attention on the closed interval with endpoints $t$ and $t'$.
 Thus, $O_I$ describes the stream of attention over the time period $I$, whereas the quotient space $o_I$ collapses any portions of this space that consist of frozen streams of attention (thus, $o_I$ describes just the the changes in attention over the time interval $I$).  
 The motivation for introducing this quotient step is to allow the possibility for a mind to hold a representation of attention over time if the attention is not shifting with time, and thereby strengthening the statement of our main result. 

 \subsection{Attention schema}
 
  A {\em formal attention schema}\footnote{The word `formal' is to emphasise a mathematical abstraction of the notion from AST.}
for $(\A, F)$  is a connected open subspace  $\A_\cS$ of $\A$ together with a map $\varphi_I$ from any quotient (fixed or moving) stream of attention $o_I$ for  $\A$  into $\A_\cS$.  We denote the collection of the maps $\varphi_I$ (over all intervals $I$ and for a fixed $\A$) by  $\varphi$.

A simple but trivial attention schema is provided by selecting some fixed point $a \in \A_{\cS}$, and letting $\varphi$ map each point in $o_I$ (for all $I$) to  $a$ in $\A_{\cS}$.  A slightly less trivial example, would be to map each $o_I$ to its own particular (distinct) point in $\A_{\cS}$; however, this still misses much of the detail of $o_I$, even when $o_t$ is frozen in the interval $I$.
 To provide a formal attention schema for $(\A, F)$ that has greater representational detail we will say that $(\A_{\cS}, \varphi)$  is {\em complete} if $\varphi_{I}$ is injective for all time intervals $I$ of length up to some given (positive) duration $T$, otherwise, we say that $(\A_{\cS}, \varphi)$  is {\em incomplete} (the latter means that different streams of attention (up to duration $T$) cannot be distinguished, as they map to the same state in the attention schema). 

Note that we are not necessarily assuming (in a given mind) that each stream of attention $o_I$ is continually being encoded in $\A_{\cS}$.  Rather, completeness simply means that the attention schema is sufficiently detailed that it has the capability to do  so over any (short but positive-length) interval,  and in such a way that the encoding is faithful (i.e. injective).

\subsection{Moving streams of attention}
In  contrast  to a frozen stream of attention, we say that $o_t$ as defined by (\ref{Feq})  is a 
 {\em moving stream of attention},  as $t$ varies over an interval $I$, provided that the following two conditions both hold for each $t \in I$:
 \begin{itemize}
 \item[(m-i)] If $o_t =o_{t'}$ for $t'\in I$, then $t=t'$.
 \item[(m-ii)] $o_t$ can be written as  $\{g(t,x): x\in o\}$, where $o$ is an open set in $\A$ and $g: I \times o \rightarrow \A$ is a map for which $x \mapsto g(t, x)$ is injective. %% \footnote{$g(t,o)$ denotes the set $\{g(t,x): x\in o\}$ (i.e. the image of $o$ under the map $x \mapsto g(t, x)$).}
 %$o_t = \{g(t, x): x\in o\}$,  where $o$ is an open set in $\A$ and  $g: I \times o \rightarrow \A$ is continuous with $g(t, *)$ injective for each $t \in I$.
 % \item[(m-ii)]  $o_t = \{g(t, x): x\in o\}$,  where $o$ is an open set in $\A$ and  $g: I \times o \rightarrow \A$ is continuous with $g(t, *)$ injective for each $t \in I$.
 \end{itemize}
 In words, a moving stream of attention is one that moves continuously over a period of time  (e.g. across $\A$, or by becoming larger or smaller) without either stalling or returning to a previous state (i.e. violating (m-i)) or suddenly jumping to a different region of $\A$ (i.e. violating (m-ii)).   Fig.~\ref{fig1} illustrates these notions in a simple and low-dimensional setting.

An  example of a moving stream of attention would be the continuous changes in attention from observing the mouse move towards the apple in Fig.~\ref{fig1}.
 Notice that although $o_t$ in Fig.~\ref{fig1} contains moving streams of attention, $o_t$  is not a moving stream of attention over the entire interval from  $t_2$ to $t_3$. To see why, observe that Condition (m-ii) implies that $o_t$ and $o_{t'}$ are homeomorphic (i.e. topologically equivalent)  for any pair $t,  t' \in I$, since they are both isotopic embeddings of $o$ into $\A$; in particular, $o_t$ is connected if and only if $o_{t'}$ is also connected (however, $o_t$ is disconnected at $t_2$ and connected at $t_3$).

It may be tempting to suppose  that if $O_I$ is not a frozen stream of attention, then it must contain a moving stream of attention on some sub-interval $I'$ of strictly positive length.  
However, it is easy to construct counterexamples, even when $o_t$ is described by Eqn.~(\ref{Feq}), as we will formally demonstrate in the next section.

\bigskip

\section{Formal result}

In the following result, recall that an injective map refers to any continuous function that is one-to-one (i.e. no two elements are mapped to the same point).

 \begin{proposition}
 \label{mainthm}
 \mbox{}
 Suppose that $(\A_{\cS}, \varphi)$ is a formal attention schema for $(\A, F)$ and $I$ is an interval of time  that has positive length.  The following then hold.
 \begin{itemize}
% If $O_I$ is a frozen stream of attention, then $o_I$ can be  injectively mapped into  $\A_{\cS}$.
  \item[(i)]
   If $O_I$ contains a moving stream of attention (on a subinterval of $I$ that has strictly positive length), then $o_I$ cannot be mapped injectively into $\A_\cS$.
In particular, a complete attention schema is not possible in the presence of moving streams of attention.
 \item[(ii)] 
   For any sequence $o_1, o_2, \ldots, o_k$ of open sets in $\A$ (with no bound on $k$), there is a (continuous) attentional activity function $F=F(t,x)$ for which $O_I$ contains consecutive frozen streams of attention consisting of $o_1, o_2, \ldots$, and for which  $o_I$ can be mapped injectively into $\A_{\cS}$.
 \end{itemize}
\end{proposition}

The proof of Proposition~\ref{mainthm} is given in the Appendix.  Here, we make two remarks.  Firstly, although Part (ii) suggests a way in which a stream of attention is moving over the interval $I$,  the changes are not continuous;  instead, they proceed via a sequence of `jumps' (each of which may be small, though some (or all) could also be large).   Second, if we were to allow part or all of the formal attention schema $\A_{\cS}$ to inhabit a space of higher dimension than the rest of $\A$ (i.e. to be an open bounded subspace of $\RR^{N}$ for $N>n$), then moving streams of attention (not involving $\A_\cS$) could now be mapped injectively into $\A_{\cS}$.    
However, this does still not allow for a complete formal attention schema, since attention could now move across this higher-dimensional part of $\A_{\cS}$  (i.e. a moving stream of attention across the attention schema itself). 
%That is,  the proof of Proposition~\ref{mainthm}(i) implies the following:
%\begin{corollary}
%Suppose that $\A$ is extended so that (some or all of) $\A_{\cS}$ is replaced by an open subspace $\A'_{\cS}$ of $\RR^N$, where $N$ is greater than the dimension of the rest of $\A$. Then a complete attention schema is not possible in the presence of moving streams of attention across $\A'_{\cS}$. 
%\end{corollary}

\section{Concluding comments}
Proposition~\ref{mainthm} required a number of assumptions, and we comment briefly on the impact of relaxing or removing them. Firstly, if $\A$ is allowed to have cells of infinite or arbitrarily high dimension (e.g. to be a subspace of the countably infinite union of the finite-dimensional spaces $\RR^{i}$ for $i=1,2,3\ldots$), then Proposition~\ref{mainthm}(i)  need not hold. Similarly, if $\A$ is allowed to vary with $t$, so that cells of increasing dimension can continue to be generated as time advances, then Proposition~\ref{mainthm}(i) may again not hold (at any given time).  However, if one accepts that a mind is entirely dependent on a nervous system (and external stimuli), which, though extremely complex, are nevertheless of bounded dimensions, the assumption of a finite (albeit large) dimensional topological space seems reasonable. 

Our theoretical result offers some insight into the generic properties of such models such as AST.  Other mathematical approaches to consciousness have been explored in a number of recent papers  (see e.g. \cite{ale, mag, moy, rud, sig, vel}); however, the approach here is quite different.  The use of topology has proven useful in identifying generic properties for a range of other processes and systems  in applied science (see e.g. \cite{ghi}).  Indeed, an  applied branch of algebraic topology --- persistent homology --- has led to some important data-driven insights into  the structural topology of brain processes, including the neural correlates of consciousness  \citep{exp, pet}; again, the approach here (which focuses on AST) has no direct relationship to this earlier work. 

\section{Declaration of Competing Interest}
The author declares that he has no known competing financial interests or personal relationships that could have appeared to influence the work reported in this paper.

\section{Acknowledgements} I wish to thank the reviewers for several helpful comments on an earlier version of this manuscript, and Jasmine Peate-Garratt for sketching some images used in Fig. 1.

%\section{References}

\bibliographystyle{model2-names}
\bibliography{JTB_cons_revision2.bib}

%\newpage

\section*{Appendix: Proof of Proposition~\ref{mainthm}}

\noindent{\em Part (i):}  Suppose that  $O_{I'}$ is a moving stream of attention  (for some subinterval $I'$ of $I$ of strictly positive length). Recall that $O_{I'} =  \{(t, x): t\in I', x\in o_t\} \subseteq I' \times \A$, and $o_{I'}$ is the quotient space $O_{I'}/\sim$. 
By Condition (m-i),  no two points of $O_{I'}$ are equivalent under the relation $\sim$,  so $o_{I'}$ is homeomorphic to $O_{I'}$. 

 Now consider the function $g$ that is the subject of Condition (m-ii) (using $I'$ in place of $I$),   let  $$Y=\{(t, x): t \in I', x\in o\},$$ and let
$$H: Y \rightarrow O_{I'}$$ be  defined by $H(t,x) = (t, g(t,x))$.  

We claim that $H$ is an injective map. To see this, observe that the continuity of $H$ follows from the continuity of $g$, whereas  for  injectivity, the equality $H(t,x)=H(t',x')$ implies that  $(t, g(t,x))=(t', g(t', x'))$ and so $t=t'$, and hence $g(t, x)=g(t, x')$, thus  $x=x'$ by the injectivity of the function $x \mapsto g(t, x)$ for each fixed $t$, from Condition (m-ii)). 

The set $Y$ contains an open cell $e_{n+1}$ of dimension $n+1$ (since $I'$ contains an open 1-cell, and $o$ contains an open $n$-cell, and the resulting  product space of these two cells  is an open $(n+1)$-cell). 

Let us now suppose that $\varphi_I$ is injective (we will show that this is impossible by deriving a contradiction). We then have the following composition of injective maps:
\begin{equation}
\label{long}
 e_{n+1}\xrightarrow{i}  Y  \xrightarrow{H} O_{I'}  \xrightarrow{\epsilon} o_I \xrightarrow{\varphi_I}  \A_\cS \xrightarrow{i'} \RR^n,
 \end{equation}
where $i$ and $i'$ are subspace inclusion maps, and $\epsilon: O_{I'} \rightarrow o_I$ maps $x$ to $[x]$ where $[*]$ refers to the equivalence class under $\sim$ ($\epsilon$ is injective by Condition (m-i)). 

We now apply a classical result from algebraic topology: the Invariance of Domain theorem  (e.g. \cite{hat}). One form of this theorem states that if $f: o \rightarrow o'$ is an injective map from an open $m$-cell to  $\RR^n$, then $m \leq n$.
Applying this with $m=n+1$, and noting that the composition of injective maps in (\ref{long}) gives an injective map $f= i' \circ \varphi_I \circ \epsilon  \circ H \circ i$ from  $e_{n+1}$ into $\RR^n$ which results in the contradiction $n+1 \leq n$.  Thus $\varphi_I$ cannot be injective, as claimed.

\bigskip

{\em Part (ii):}  We first show that if $o$ is an open set in $\A$ and $\delta>0$ is a threshold, then there is a continuous function $\psi_o: \A \rightarrow \RR^{\geq 0}$  for which $o=\{x\in A: \psi_o(x)> 0\}$.  To see why this holds, let $\partial(o)$ denote the boundary (topological frontier) of $o$ in $\A$ (since $o$ is open in $\A$, this is the set of points in $\A$ that are in the closure of $o$ but not in $o$), and let $D(x,  \partial(o)) = \inf \{d(x,y): y \in \partial(o))$, where $d$ is any standard metric on $\RR^n$ (e.g. the Euclidean metric). 
Since $\partial(o)$ is closed and bounded, and therefore compact (by the Heine-Borel theorem), it follows that $D(x, \partial(o))$ is realised by at least one point $y \in \partial(o)$.  Let:
$$\psi_o(x) = \begin{cases}
D(x, \partial(o)), & \mbox{ if $x \in o$};\\
0, &\mbox{ if $x \in \A-o$}.
\end{cases}
$$
The function $\psi_o$ is well-defined, continuous, and $\psi_o(x)> 0$ if $x \in o$  and  $\psi_o(x)=0$ if $x \in \A-o$. In particular, 
$o=\{x\in \A: \psi_o(x)>0\}$, as claimed.

Without loss of generality, we may assume that $I$ is the open interval $(0,1)$.  Write $I$ as a sequence of open intervals of $I$ of equal length interspersed with singleton elements of $I$ as follows:
$$(0, \frac{1}{k}), [\frac{1}{k}], (\frac{1}{k}, \frac{2}{k}), [\frac{2}{k}],  ( \frac{2}{k}, \frac{3}{k}) \cdots [\frac{k-1}{k}], (\frac{k-1}{k}, 1).$$
For $j=1, \ldots, k$, let $I_j$ denote the open interval $(\frac{j-1}{k}, \frac{j}{k})$, and for $t \in I_j$ let 
 $$\gamma_j(t) =\begin{cases}
 \min\left\{t-\frac{j-1}{k}, \frac{j}{k}-t\right\}, & \mbox{ if $t \in I_j$};\\
 0, & \mbox{ otherwise.}
 \end{cases}
 $$
 Thus $\gamma_j(t)$ is continuous on $I$, and is non-zero and strictly positive precisely on $I_j$.
We now define an attentional activity function $F: (0,1) \times \A \rightarrow \RR^{\geq 0}$ by setting:
$$F(t, x) =  \delta + \sum_{j=1}^k \gamma_j(t) \cdot \psi_{o_j}(x).$$
The function $F$ is then well-defined,  continuous in both variables,  and $F$ has the property that $o_t$ defined by Eqn.~(\ref{Feq}) coincides with $o_j$ throughout  $I_j$ for $j=1, \ldots, k$ (at the singleton element $t_j=\frac{j}{k}$ that appears between $I_{j}$ and $I_{j+1}$ for $j=1, \ldots k-1$ we have:
$o_{t_j}= \emptyset$). 
Thus  $O_I$ contains frozen versions of $o_1, o_2, \ldots, o_k$ (in this order) and $o_I$ consists of $k$ connected components, namely disjoint copies of  $o_{1}, o_{2}, \ldots, o_{k}$.
Since each set $o_j$  is an open set in $\A$ (and therefore an open set of $\RR^n$), there is an injective map of each set $o_j$  into any given open $n$-cell $e_n$ contained in $\A_{\cS}$;  therefore, there is an injective map of  $o_I$ into $e_n$ as well (by selecting $k$ disjoint open $n$-cells within $e_n$ to map each $o_j$ set into).   

Note that our construction provides just one way of guaranteeing the existence of a (continuous) attentional activity function $F$ satisfying the property required;  many other possible choices for $F$ are possible.
\hfill$\Box$

\end{document}